\documentclass[12pt]{article}
\usepackage{amsmath}

\newcommand{\be}{\begin{equation}}
\newcommand{\ee}{\end{equation}}
\usepackage{graphicx}

\def\eqn#1{$$#1$$}
\def\eqneqn#1{\begin{multline*}#1\end{multline*}}

\def\eqnn#1#2{
	\begin{equation}#2\label{#1}
	\end{equation}}
\def\eln#1{\begin{align*}#1\end{align*}}
\def\elnn#1#2{\begin{align}\begin{split}#2\label{#1}
	\end{split}\end{align}}

\begin{document}
\baselineskip=24 pt

\begin{center}

{\large {\bf Closed timelike curves re-examined}}

{\tiny(\today)}

\end{center}

\vskip1.5truecm 

\begin{center}
  F. I. Cooperstock and S. Tieu \\
{\small \it Department of Physics and Astronomy, University 
of Victoria} \\
{\small \it P.O. Box 3055, Victoria, B.C. V8W 3P6 (Canada)}\\

{\small \it e-mail addresses: cooperstock@phys.uvic.ca, stieu@uvic.ca}
\end{center} 

\begin{abstract} 
Examples are given of the creation of closed timelike curves by choices of 
coordinate identifications. Following G\"odel's prescription, it is seen that 
flat spacetime can produce closed timelike curves with structure similar to that of G\"odel. In this context, coordinate 
identifications rather than exotic gravitational effects of general relativity 
are shown to be the source of closed timelike curves. Removing the periodic 
time coordinate restriction, 
the modified G\"odel family of curves is expressed in a form that retains
the timelike and spacelike character of the coordinates. With these 
coordinates,
the nature of the timelike curves is clarified. A helicoidal surface unifies 
the families of timelike, spacelike and null curves.
In all of these, it is seen that as in ordinary flat spacetime, periodicity in
the spatial position
does not naturally carry over into closure in time. Thus, the original 
source of serious scientific speculation regarding time machines is
seen to be misconceived. 
\end{abstract} 
\vspace*{1truecm}

\begin{center}
\mbox{ }\\
\end{center}

\section{Introduction}

While the notion of time travel has excited the imagination of the general 
public for decades if not centuries, the possibility that there was a
potential for its realization as a part of serious scientific investigation
is usually attributed to the discovery of closed timelike curves (CTC's)
in the G\"odel universe of general relativity\cite{Godel}.
A CTC is defined as a timelike future-directed curve
(i.e. always evolving in time within the future light
cone), 
reuniting with a spacetime point of its earlier history and hence recycling 
endlessly. The general view has been and continues to the present that the 
scope of exotic gravitational phenomena via general relativity is the source 
of G\"odel's closed timelike curves. In this paper, we show that the essential source 
is actually a rather unnatural choice of identifying spacetime points and that 
the natural choice does \textit{not} lead to CTC's.

The G\"odel spacetime, describing a type of 
rotating universe with no expansion, is a particular example of the generic 
class given by\cite{Bonnor}
\eqnn{Eq1Bonnor}{
	ds^2 = -f^{-1} [ e^\nu( dz^2+dr^2) +r^2 d\phi^2]+f(d\bar{t}-wd\phi)^2
}
where $f$, $\nu$ and $w$ are functions of $r$ and $z$ with the
coordinates having the ranges
\eqnn{Line5ofBonnor}{
	-\infty<z<\infty,\quad
	0\le r,\quad 
	0\le \phi\le 2\pi,\quad
	-\infty<\bar{t}<\infty
}
and $\phi = 0$ and $\phi = 2\pi$ being identified. The standard argument
is the following: the metric component
\eqnn{Line7ofBonnor}{
	g_{\phi\phi} = -f^{-1} ( r^2 -f^2 w^2 )
}
changes sign at the point where $f^2w^2 = r^2$ and hence $\phi$ becomes a 
timelike coordinate for 
\eqnn{EqA}{
 	f^2w^2>r^2.
}
In this case, the spacetime curve 
\eqnn{Line11ofBonnor}{
	\bar{t}=\bar{t}_0,\quad r=r_0,\quad \phi=\phi,\quad  z=z_0
}
with $z_0$, $r_0$, $\bar{t}_0$ being constants has been created as a CTC
as a result of the now-timelike coordinate $\phi$ having $\phi=0$ and $\phi = 
2\pi$ still being identified as was the case when $\phi$ was spacelike.

There are essential problems with such an approach: while the 
interpretation of the nature of the curve for 
\eqnn{EqB}{
	f^2w^2<r^2
}
as being one of a closed spacelike curve on a constant time slice is 
unassailable, in case (\ref{EqA}), the metric has {\em two} timelike
coordinates $\bar{t}$ and $\phi$. One coordinate $\bar{t}$ is held fixed
while the other coordinate $\phi$ is allowed to
run.  In spite of the fact that there is nothing inherently wrong in 
coordinatizing a spacetime with more than one 
timelike coordinate (Synge\cite{Synge} provides an example of doing
so with four timelike coordinates\footnote{
	Having 4 timelike coordinates refers to having the diagonal terms
	of $[g_{ij}]$ all being positive. If it is a physical spacetime,
	the signature, derived from its eigenvalues, will still have the
	signs $(+ - - -)$.}),
it is not conducive to clarity to have a timelike coordinate held fixed in the 
description of a timelike curve.
In what follows, we explore the ramifications of imposing a periodicity upon 
the timelike coordinate $\phi$.

It has often been remarked that one can artificially create a CTC in $1+1$ 
Minkowski spacetime
with the usual coordinate ranges and metric
\eqnn{EqC}{
	ds^2=dt^2-dx^2
}
provided one imposes the condition that the points $(t,x)$ and $(t+t_0,x)$  
($t_0$ is a constant) are identified. With this periodic identification in the 
time coordinate, the essential nature of the spacetime is altered from its 
standard form. This kind of identification plays a key role in what follows.

As a 
second example, we consider
flat $3+1$ spacetime in cylindrical polar coordinates
\eqnn{EqF}{
	ds^2= dt^2 - dr^2 -r^2d{\phi}^2-dz^2
}
with the standard coordinate ranges and where $\phi=0$ and $\phi = 2\pi$ are 
identified as usual, i.e.
\eqnn{oldid}{
        (t,r,0,z)=(t,r,2\pi,z)
}
We retain the identification in $\phi$ for $0$ and $2\pi$ as we effect the 
transformation 
\eqnn{EqG}{
	\bar{t} = t + a\phi, \quad \bar{\phi}= \phi, \quad \bar{r}=r, 
       \quad \bar{z}=z
}
where $a$ is a constant.
\footnote{\label{fn:recipe}
        If we wish to preserve the structure of the spacetime,
	we must apply this transformation to (\ref{oldid})
	and re-express it in the new coordinates.
        To do so, first define $P_1=(t_1,r_1,0,z_1)$
        and $P_2=(t_1,r_1,2\pi,z_1)$.
        Then transform $P_2$ into
        $\bar{P}_2=(\bar{t}_2,r_2,\phi_2,z_2)$
        and using $t_1 = \bar{t}_1 - a\phi_1$, write the
        transformed $\bar{P}_2$ in terms of $\bar{t}_1$.
        Finally, apply $\bar{P}_1=\bar{P}_2$ leading to the result of
        $(\bar{t},r,0,z)=(\bar{t}+2\pi a,r,2\pi,z)$.
        It should be noted that,
        while this identification appears somewhat
        unusual in the new coordinates, it
        will not induce any singularity in the
        curvature similar to the vertex of a cone.
	Thus this particular system differs from a
        circumnavigated cosmic string.
}
The metric becomes
\eqnn{EqH}{
	ds^2= d\bar{t}^2- dr^2 - 2ad\bar{t}d\phi -(r^2-a^2)d{\phi}^2 -dz^2
}
which is precisely of the type (\ref{Eq1Bonnor}) but with constant values
globally for $f$, $w$ and $\nu$. 

\begin{figure}
\begin{center}
\includegraphics[width=6in]{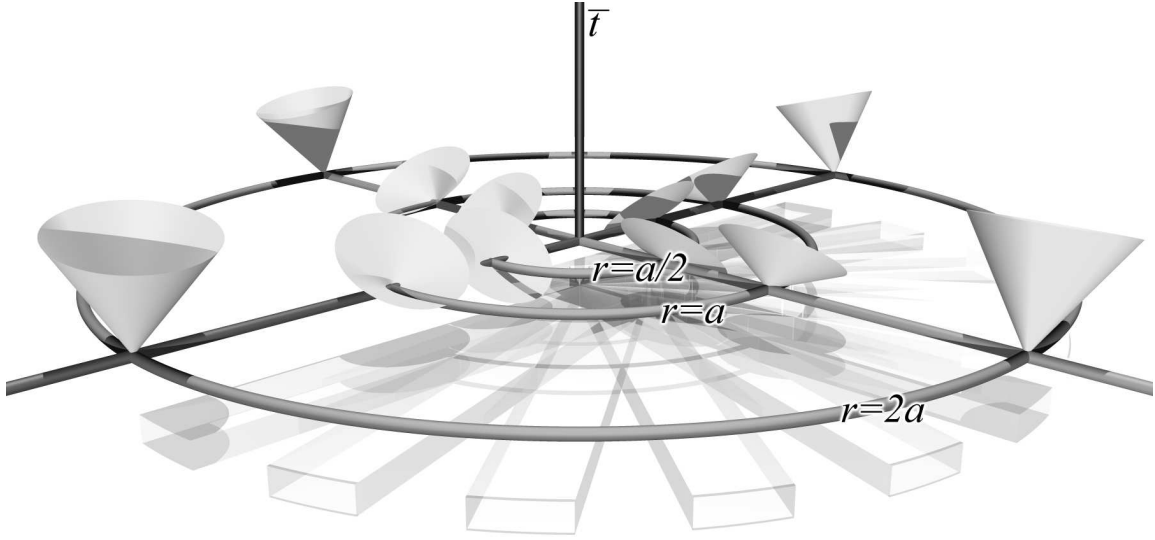}
\end{center}
\caption{
	In the $(\bar{t},\phi)$ coordinates, the tipping light cones
        produce a CTC for $r<a$. The boxes at the bottom
        follow the curves for constant $\bar{t}$.
}
\label{fig:godel1}
\end{figure}

Here we shall follow the standard argument to conclude that
our spacetime, with metric (\ref{EqH}), contains closed timelike $\phi$ curves
$(\bar{t}_0,r_0,\phi,z_0)$ for $r_0^2<a^2$.
First, the indication of the timelike character is the positive sign of the
$g_{\phi\phi}$ component of (\ref{EqH}).
Second, the indication of closure follows from the \textit{imposed} closure
characteristic of the $\phi$ coordinate,
\eqnn{secondid}{
	(\bar{t},r,0,z) = (\bar{t},r,2\pi,z).
}
Note that this identification is not equivalent to (\ref{oldid}).
By the analysis of the lightcones, we develop the standard
figure \ref{fig:godel1} depicting the transition from closed spacelike to null 
to timelike curves.

\begin{figure}
\begin{center}
\includegraphics[height=3in]{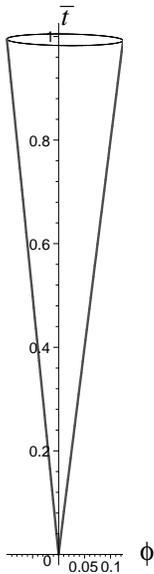}
\end{center}
\caption{The lightcone 
becomes narrow for $r\gg a$.}
\label{fig:lightconefar}
\end{figure}

\begin{figure}
\begin{center}
\includegraphics[width=3in]{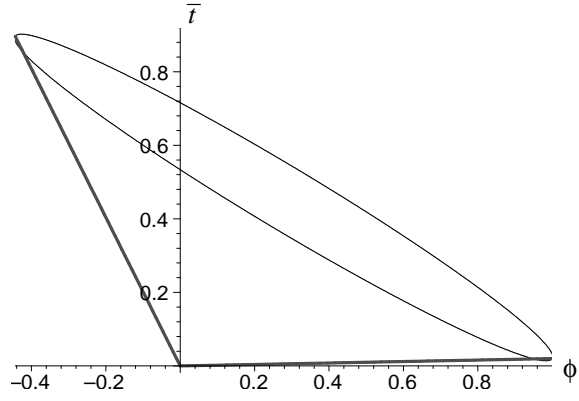}
\end{center}
\caption{The critical value of  $r\rightarrow a$ results in the lightcone 
touching the $\phi$ axis.}
\label{fig:lightconecrit}
\end{figure}

\begin{figure}
\begin{center}
\includegraphics[width=3in]{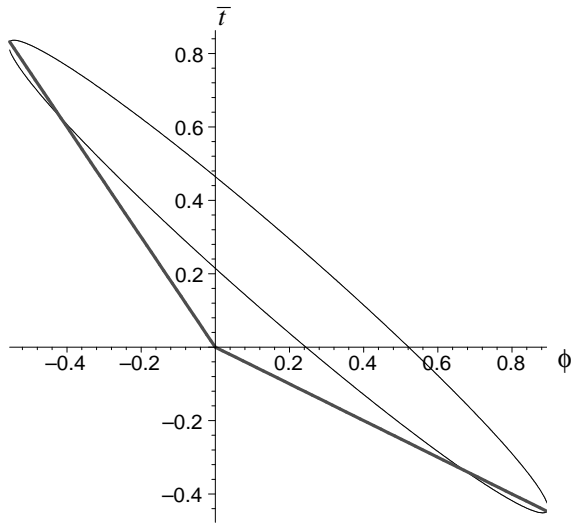}
\end{center}
\caption{The lightcone structure of $r=a/2$ where the $\phi$ direction is 
timelike.}
\label{fig:lightconenear}
\end{figure}

In the examination of the lightcone structure, we will see in what
follows that these $\phi$-curves are indeed spacelike for
certain values of $r$ and timelike for others.
In the original coordinates, the null vectors in the $\phi$ direction
have ``velocities''\footnote{Here, the velocities are measured in
radians per unit time.}
\eqnn{EqNullOld}{
	\left( \frac{d\phi}{dt} \right)_{\text{null}}
	= \pm \frac{1}{r}
} 
whereas in the new coordinates, they are
\eqnn{EqNullNew}{
	\left( \frac{d\phi}{d\bar{t}} \right)_{\text{null}}
	= \frac{1}{a\pm r}.
}
As shown in figure \ref{fig:lightconefar},
in the limit as $r\rightarrow\infty$, the lightcone (\ref{EqNullNew})
becomes very narrow (similar to the case in the original coordinates
(\ref{EqNullOld})).
At the critical radius $r\rightarrow a^+$,
the lightcone dips and touches the $\phi$ axis\footnote{
	The ${\mathcal M}_4 \longrightarrow (\bar{t},r,\phi,z)$
	chart does not have any ``$\phi$ axis.''
	The $\phi$ axis refers to the ${\bf e}_\phi$ direction in 
	$T_p({\mathcal M}_4)$ for a particular point $p$. At each point $p$ in 
the 4-dimensional
(flat) spacetime manifold ${\mathcal M}_4$, there is a tangent vector
space $T_p({\mathcal M}_4)$, where all vectors
${\bf u}=u^i {\bf e}_i$ reside.  However, only the 
${\bf e}_\phi$ and ${\bf e}_t$ directions are of interest
because our curve will not have
any ${\bf e}_r$ or ${\bf e}_z$ components.
The lightcone is defined as a set of all
vectors ${\bf u}=u^i {\bf e}_i \in T_p({\mathcal M}_4)$
such that ${\bf g}({\bf u},{\bf u})=g_{ij}u^i u^j =0$.
In the metric (\ref{EqF}) and (\ref{EqH}), it is obvious that a
single-component vector ${\bf u} = u^\phi {\bf e}_\phi$ cannot be
a null vector. Thus, when we refer to null vectors in the ``$\phi$ direction''
(as opposed to the $r$ or $z$ directions), we are referring to all
vectors in $T_p({\mathcal M}_4)$ 
having both ${\bf e}_\phi$ and ${\bf e}_t$ components.
It is useful to compare our curve with the lightcone at each point.
It should be emphasized that in such a comparison, 
we \textit{first} pick a particular point $p$ and \textit{then} work
within the tangent vector space $T_p({\mathcal M}_4)$. In this vector
space it is sensible to speak of the ``$\phi$ axis'' whereas the 
${\mathcal M}_4$ manifold does not have any ``$\phi$ axis.''
} 
as seen in
figure \ref{fig:lightconecrit}. Figure \ref{fig:lightconenear}
illustrates the structure for $r_0<a$; the $\phi$ curve is enclosed within 
the lightcone. Together, these can be combined to construct a
diagram similar to that which displays the curves of the G\"odel universe as 
shown in the standard texts (see for example \cite{hawking}).
Figure \ref{fig:godel1} would indicate that for $r_0<a$,
there are closed timelike curves.

Consider the facts which imply that the $\phi$-curve is a closed
timelike curve for a fixed $r_0<a$.
The curve is always timelike, and hence the proper time flows
monotonically and never becomes imaginary, i.e. the curve does not reverse
and proceed into the past lightcone.
If we transform the ``cylindrical coordinates''
$(\bar{t},r,\phi,z)$ into a more familiar ``cartesian coordinates''
$(\bar{t},\bar{x},\bar{y},\bar{z})$,
we find that the $\phi$ curve follows the trajectory
\eln{
	\bar{t} &= \bar{t}_0 \\
	\bar{x} &= r_0 \cos \phi \\
	\bar{y} &= r_0 \sin \phi \\
	\bar{z} &= z_0 \\
	ds^2 &> 0 \quad \text{ (time-like) }\forall \phi \in [0,2\pi]
}
and this timelike curve returns to the original location in spacetime as a CTC.

However, we recall that the original spacetime, with metric (\ref{EqF}) and 
standard coordinate ranges and identifications, is simply ordinary flat
spacetime. The metric (\ref{EqH}) was derived simply from a coordinate
transformation.
The essential element that led to the CTC in this flat space was the continued 
demand that $\phi$ exhibit closure even when it became a timelike coordinate.

\begin{figure}
\begin{center}
\includegraphics[width=6in]{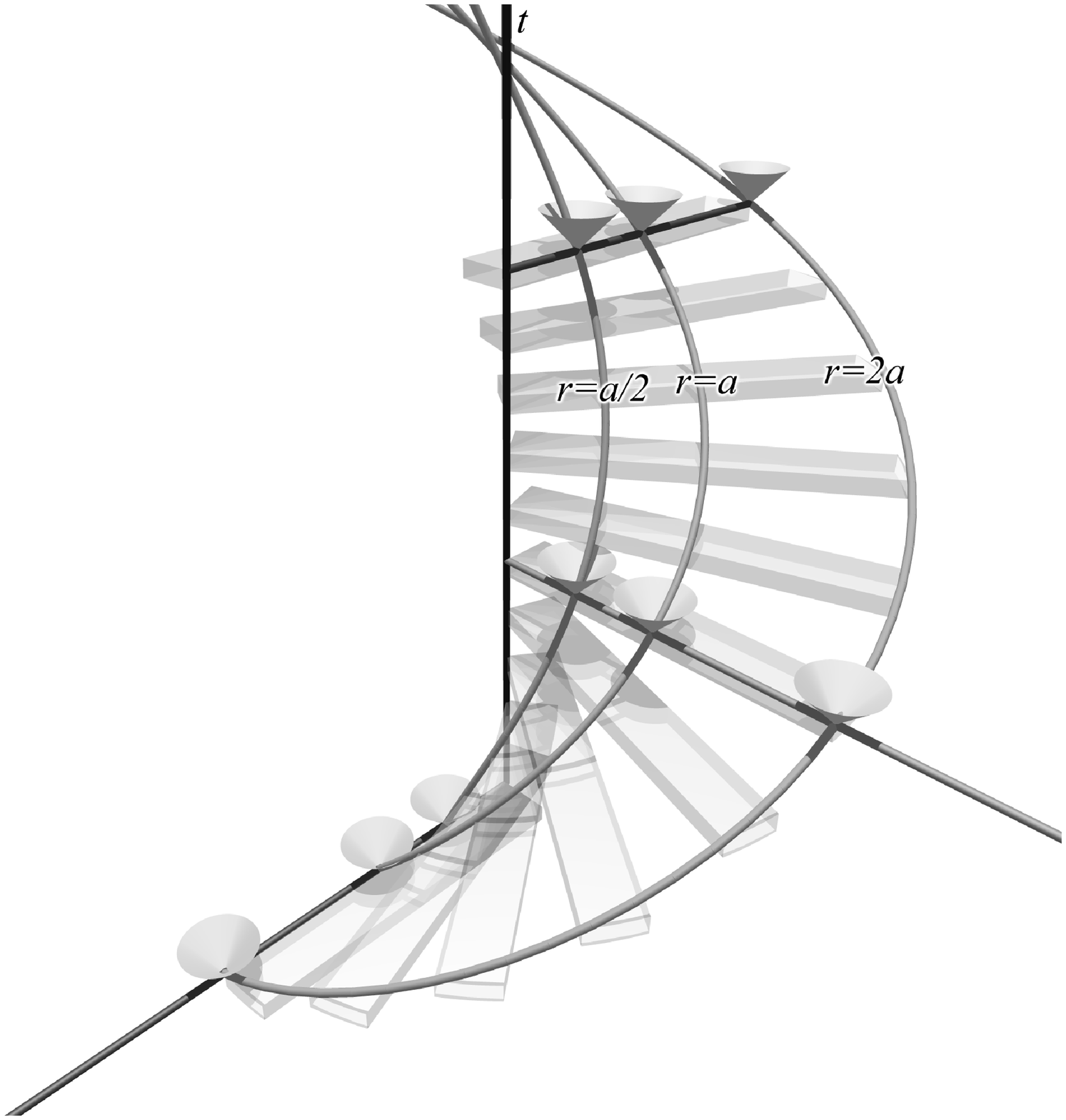}
\end{center}
\caption{
	Again, the boxes are used as visual aids to illustrate the evolution
	of the curves.
        By contrast with the previous figure, the boxes here are at
        constant $t$.
In the $(t,\phi)$
        coordinate system, the spacelike, null and timelike curve are seen as 
a unified family of curves advancing monotonically in time t. 
        Evolving curves never close in terms of $t$ and hence there are
        no CTC's with the periodic time restriction removed.  Here, the fixed 
$\bar{t}=\bar{t}_0$ surface is
        actually  helicoidal.
}
\label{fig:godel2}
\end{figure}

To illustrate a more natural choice of identification for these curves, we 
transform the light cones 
of figure \ref{fig:godel1} back into the original fiducial $(t,r,\phi,z)$
coordinates. This is illustrated in figure \ref{fig:godel2}.
One can see why these lightcones in figure \ref{fig:godel1}
appear to tilt in terms of the $(\bar{t},\phi)$ coordinates
as $r$ varies.  The curves $t+a\phi=\bar{t}_0$, $r=r_0$, $z=z_0$
being helices, are inside and outside the lightcone for
$r_0<a$ and $r_0>a$, respectively. If we do not continue to impose closure in 
$\phi$ when it becomes a timelike coordinate, the CTC characteristic is absent.

\begin{figure}
\begin{center}
\includegraphics[height=3in]{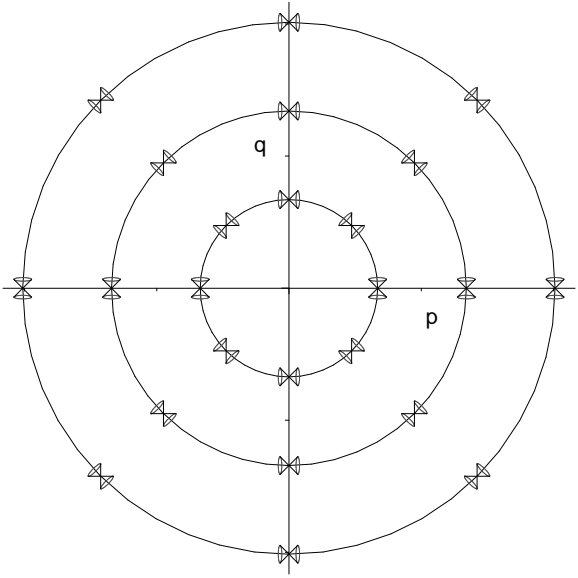}
\end{center}
\caption{The lightcones in $(p,q)$ coordinates.}
\label{fig:morelightcones}
\end{figure}

In  G\"odel's \cite{Godel} spacetime, the metric 
\eqnn{godelfirstform}{
	ds^2 = a^2
	\left( d\bar{t}^2-d\bar{r}^2
	+\frac{1}{2} e^{2\bar{r}} d\bar{\phi}^2
	+ 2e^{\bar{r}} d\bar{t} d\bar{\phi}
	-d\bar{z}^2
	\right)
}
is expressed with timelike coordinates $\bar{t}$,$\bar{\phi}$ globally.
To present this in a physically desirable globally explicit 3+1 form, the transformation
\eln{
	\bar{t} &= t+\frac{r\phi}{2}\left(1-\ln r\right)+\frac{1}{2} \ln r\\
	\bar{r} &= r\phi \\
	\bar{\phi} &= -\frac{1}{2} e^{-r\phi} \ln r \\
	\bar{z} &= z
}
is applied.
The metric becomes
\eqneqn{
	\frac{ds^2}{a^2}
	=dt^2
	-\left[\phi^2+\frac{1}{8r^2}\left(r\phi\ln r -1\right)^2\right]dr^2
	-\left[\frac{3}{4}r^2 + \frac{1}{8}(r\ln r )^2\right]d\phi^2
	-dz^2 \\
	-\frac{1}{4}\left( 8r\phi +r\phi(\ln r)^2 -\ln r \right)dr d\phi
	+r dt d\phi.
}
The price to pay for achieving this more desirable form is the introduction of $\phi$ dependence in the metric, a phenomenon that is familiar from other situations in general relativity.
The identification
\eqn{
	(\bar{t},\bar{r},0,\bar{z})
	=(\bar{t},\bar{r},2\pi,\bar{z})
}
is transformed, following the procedure as in footnote \ref{fn:recipe}, to
\eqn{
	(t,1,\phi,z)
	=(t+2\pi(1-\phi)e^\phi,e^{-4\pi e^\phi},\phi e^{4\pi e^\phi},z)
}
and in this form, there is no suggestion of any identification of spacetime points.
One might object that there was no motivation to identify the
$\bar{\phi}$ end-points in the first G\"odel form (\ref{godelfirstform})
whereas in the second form \cite{Godel}
\eqn{
        ds^2 =
        4a^2
	\left(
	d{T}^2-dR^2+(\sinh^4 R-\sinh^2 R)\,d\varPhi^2
        +2\sqrt{2} \sinh^2 R \, dT d\varPhi
        -dZ^2
	\right)
}
there is a 
motivation to identify $\varPhi=0$
and $\varPhi=2\pi$ because of the transformation
\eln{
        e^{\bar{r}}
		&= \cosh 2R + \cos \varPhi \sinh 2R \\
        \bar\phi e^{\bar{r}}
		&= \sqrt{2}\sin{\varPhi} \sinh 2R \\
        \tan\left(\frac{\varPhi}{2}+\frac{\bar{t}-2T}{2\sqrt{2}}\right)
        	&=  e^{-2R} \tan \frac{\varPhi}{2},
}
i.e. $\varPhi=0$ and $\varPhi=2\pi$ are mapped to the same
point due to the $\sin$, $\cos$ and $\tan$ of $\varPhi$ 
terms in the transformation. Because of this, one might argue that
the $\varPhi$-curve is \textit{naturally} closed (as well as being timelike).
To counter this argument, consider a simple Minkowski spacetime $(t,x)$
mapped to $(p,q)$ using
\eln{
	p&=x\cos t \\
	q&=x\sin t
}
The metric $ds^2 = dt^2-dx^2$ becomes
\eln{
	ds^2 =
	\left( \frac{q^2}{(p^2+q^2)^2} - \frac{p^2}{p^2+q^2} \right)
	dp^2
	+
	&\left( \frac{p^2}{(p^2+q^2)^2} - \frac{q^2}{p^2+q^2} \right)
	dq^2 \\
	-
	2pq
	&\left( \frac{p^2 +q^2 +1}{(p^2+q^2)^2}\right) dp dq.
}
The light cones are shown in figure \ref{fig:morelightcones}.
The Jacobian of the transformation vanishes only for $x$=$0$.
Consider the curve $p=\cos \tau$, $q=\sin \tau$ which is time-like and 
\textit{naturally} closed. 
Clearly, this timelike curve is simply a segment from $(t,x)=(0,1)$ to $(2\pi,1)$
and it would be quite \textit{unnatural} to identify these two end-points.

Returning to the case of the G\"odel spacetime, one might object that in this process, 
G\"odel's CTC has been artificially removed by making different identifications. However, it is more 
persuasive to argue that it is the identification in the original G\"odel 
spacetime that is the artificial one.
This is reminiscent of our first example
where one could choose to have or not have a CTC, simply through the choice of 
identification.
CTC's did not manifest themselves in the metric, and gravitation via general 
relativity was not the factor that led to their realization.

We now return to (\ref{Eq1Bonnor}) and consider the transformation for
the curve where $r$, $z$ (and hence $f$, $\nu$ and $w$) are held
constant as
\elnn{EqI}{
	dt &= d\bar{t} -wd\varphi \\
	d\Phi &= \frac{w^2f-r^2f^{-1}}{2fw} d\varphi-d\bar{t}.
}
The line element for this curve is
\eqnn{EqJ}{
	ds^2= \frac{f}{(w^2f^2+r^2)^2}
	\Bigl((w^2f^2-r^2)^2dt^2
		-8f^2w^2r^2d\Phi dt
		-4f^2w^2r^2d\Phi^2\Bigr).
}
In this system, $t$ is a timelike coordinate and $\Phi$
is a spacelike coordinate regardless of whether (\ref{EqA})
or (\ref{EqB}) holds. With these coordinates, the integrity of the azimuthal 
coordinate is maintained explicitly unlike the case with the G\"odel approach.
 Ambiguities of interpretation
are removed and hence these coordinates are particularly 
valuable. With $\bar{t}$ held constant, say $\bar{t}=0$ for simplicity,
we find that the curve has the equation in parametric form
\elnn{EqK}{
	t&=-w\varphi \\
	\Phi&=\frac{(w^2f-r^2f^{-1})\varphi}{2fw}
}
with parameter $\varphi$.

With the parameter eliminated between the two equations, we see
that $\Phi$ is simply a linear function of $t$ with 
proportionality factor dependent upon the particular $(r,z)$ chosen.
At this point, it is natural to ascribe standard geometric
(in this case cylindrical polar-like) character by
identifying\footnote{
	In this more general case, the identification in $\Phi$
	is at $0$ and $2\pi(w^2f-r^2f^{-1})/(2fw)$. 
        This is seen as an indicator of a singularity at $r=0$ analogous
	to the vertex of a cone. 
        Bonnor has argued that the CTC could be interpreted
        as a torsion singularity \cite{Bonnor}.
}
the spatial points for $\Phi$ (rather than for $\varphi$).
However, there is no reason to ascribe
periodicity to $t$. The time flows monotonically without
looping as it does in conventional flat space. Thus, while the spatial points 
are
retraced $\textit{ad infinitum}$, they do
so at successively later times. They do so here as in the previous examples in figure \ref{fig:godel2}.

While the imposition of periodicity in $\phi$ follows naturally from our 
experience with the symmetry of various spatial geometries, there is no logic 
that leads us to continue to impose closure in $\phi$ when it becomes a 
timelike coordinate.  While our experience with nature leads us to deem as 
physical only those timelike curves that evolve into the \textit{future} 
lightcone, there is nothing in our experience that would have us place any 
\textit{apriori} demands for periodicity in a timelike coordinate.
If gravity were to truly cause closure in a timelike curve, it should do so without 
injecting periodic character in the timelike coordinate from the outset.

We have seen how transformations with $t$ and $\phi$ of the form 
$\bar{t}=t+a\phi$ generate
helicoidal surfaces for a fixed time $\bar{t}$ as viewed in
the $t$,$\phi$ coordinate system.
While mixing a non-cyclic variable, $t$ with a
cyclic $\phi$ usually does not create difficulties in other branches of
physics, the interpretation is more critical in general relativity.

For a proper interpretation, the nature of the coordinates should not be 
ambiguous.  When the metric is expressed with more than one timelike
coordinate, much confusion can result such as in the identification of
the endpoints of a curve when one of these coordinates is held fixed.
The solution is to transform to a system of coordinates in which the true 
nature of the curves in the spacetime becomes clarified.
In our example, unlike figure \ref{fig:godel1} which basically
forces one to require closure whether it is appropriate or not,
figure \ref{fig:godel2} allows the freedom to choose whether or not
one end of the curve is identified with the other.
This is achieved by finding coordinates which
maintain their timelike and spacelike character throughout.

In all the examples, identification of points can induce CTC's.
However, from the vantage point of figure \ref{fig:godel2}, it is seen
that there is no more logical basis to ascribe periodicity in
time here than there would be to do so in the case of repetitive
traversal of a circle in flat space.
In this figure, the spacelike, timelike and null curves are all
seen as part of the same family composing the helicoidal surface.
As seen from equations (\ref{EqJ}) and (\ref{EqK}), it would be
equally uncalled for to identify times in 
case (\ref{EqA}) as it would be in case(\ref{EqB}).
Unlike the standard approach with the G\"odel metric where there
is an apparent discontinuity in character of the possible types
of curves, the helicoidal surface unifies the families with all
three families advancing monotonically in time without repitition.

There are common elements throughout these examples:
A. The metric itself is insufficient to determine whether a CTC exists in the 
system.
B. The identification of points 
can induce CTC's in a
system whether that system is simple or complicated.

The vast body of physicists, not to mention the public at large, 
regard the notion of time travel as nothing more than a figment of 
science fiction fantasy.
The faith is placed in the entropy clock that is mono-directional
and non-repetitive.  However the time-machine concept has
re-surfaced in recent years in various guises, most recently in
connection with wormholes (see, for example Visser \cite{Visser}) and
through other exotic channels such as tachyons and colliding or
spinning cosmic strings. It is natural to wonder if this would
have occurred had it been appreciated from the outset that the
apparently more conservative route to time-travel in normal topology
such as with the G\"odel metric was not what had been claimed for it,
that the curves were closed in time simply because of an imposed periodicity 
in a timelike coordinate.

After G\"odel's work appeared, Chandrasekhar and Wright (``CW") \cite{chandra} 
showed that the G\"odel CTC's were not geodesics, thus concluding that his 
paper was in error. Stein \cite{stein} countered that G\"odel never claimed 
that his CTC's were geodesics and recently, Ozsvath and Schucking 
\cite{os}calculated the acceleration of the CTC's. There are interesting 
considerations that arise from this. In testing whether the CTC's were 
geodesics, it could be said that CW were simultaneously determining whether 
the CTC's could exist solely within the confines of the gravitational model. 
Once it is determined that acceleration is required, the need for a mechanism 
for achieving it, namely a machine, arises. While the role of machines in 
other contexts in general relativity is frequently ignorable (and generally 
ignored), this is not the case here. 
Repetition in time is the issue in the present context and the machine must be 
shown to partake of this repetition for consistency.
In \cite{os}, there is a discussion of the vast fuel requirements and 
relativistic velocities to achieve CTC's but there are further even more 
serious issues: the accelerating machine must also follow a closed timelike 
path, i.e. it must really be a ``time machine" with all of the machine's 
complex elements and interactions experiencing closure in time. While the relatively 
simple purely gravitational model of G\"odel (via the imposition of a periodic 
time coordinate rather than an effect of gravity), can produce such closure, albeit artificially, this has not 
been demonstrated to be possible, nor would we ever expect it to be possible, 
with the complexities of a machine that would change the metric, quite apart 
from considerations of entropy flow.

{\small {\bf Acknowledgments:} This work was supported in part by a grant
from the Natural Sciences and Engineering Research Council of Canada.} We are 
grateful to W.B. Bonnor for helpful discussions.

\clearpage
{\small 
\end{document}